\def\FLASH{{\sc flash}}
\def\PARAMESH{{\sc paramesh}}

\documentclass[letter]{aa} 
\usepackage{natbib}
\usepackage{graphicx}
\usepackage{txfonts}
\begin{document}
   \title{X-ray emission from dense plasma in CTTSs: Hydrodynamic modeling
of the accretion shock}

   \subtitle{}

   \author{G. G. Sacco
          \inst{1,}\inst{2}
          \and
          C. Argiroffi\inst{3,}\inst{2}
          \and
          S. Orlando\inst{2,}\inst{1}
          \and
          A. Maggio\inst{2}
          \and
          G. Peres\inst{3,}\inst{2,}\inst{1}
          \and F. Reale\inst{3,}\inst{2,}\inst{1}
          }

   \institute{Consorzio COMETA, Via S. Sofia, 64, 95123, Catania, Italy\\
         \email{sacco@astropa.inaf.it}
         \and
             INAF-Osservatorio Astronomico di Palermo, Piazza del Parlamento, 1, 90134, Palermo, Italy\\
         \and
           DSFA-Universit\`a degli Studi di Palermo, Piazza del Parlamento, 1, 90134, Palermo, Italy\\               
 }

   \date{Received 5 August 2008/Accepted 1 October 2008}

 
  \abstract
   {High spectral resolution X-ray observations of classical T Tauri
    stars (CTTSs) demonstrate the presence of plasma at $T\sim 2-3\times
    10^6$ K and $n_{\rm e}\sim 10^{11}-10^{13}$ cm$^{-3}$, unobserved
    in non-accreting stars. Stationary models suggest that this emission
    is due to shock-heated accreting material, but they do not allow
    to analyze the stability of such material and its position in the
    stellar atmosphere.}
   {We investigate the dynamics and the stability of shock-heated
    accreting material in classical T Tauri stars and the role of the
    stellar chromosphere in determining the position and the thickness
    of the shocked region.}
   {We perform 1-D hydrodynamic simulations of the impact of the
    accretion flow onto chromosphere of a CTTS, including the
    effects of gravity, radiative losses from optically thin plasma,
    thermal conduction and a well tested detailed model of the stellar
    chromosphere. Here we present the results of a simulation
    based on the parameters of the CTTS MP Mus.}
   {We find that the accretion shock generates an hot slab of
    material above the chromosphere with a maximum thickness of $1.8
    \times 10^9$ cm, density $n_{\rm e}\sim 10^{11}-10^{12}$ cm$^{-3}$,
    temperature $T\sim 3\times 10^6$ K and uniform pressure equal to
    the ram pressure of the accretion flow ($\sim 450$ dyn cm$^{-2}$). The
    base of the shocked region penetrates the chromosphere and stays where
    the ram pressure is equal to the thermal pressure. The system evolves
    with quasi-periodic instabilities of the material in the slab leading
    to cyclic disappearance and re-formation of the slab. For an accretion
    rate of $\sim 10^{-10} M_{\sun}$ yr$^{-1}$, the shocked region
    emits a time-averaged X-ray luminosity $L_{\rm X}\approx 7\times
    10^{29}$ erg s$^{-1}$, which is comparable to the X-ray luminosity
    observed in CTTSs of the same mass. Furthermore, the X-ray spectrum
    synthesized from the simulation matches in detail all the main
    features of the \ion{O}{viii} and \ion{O}{vii} lines of the star MP Mus.}
   {}


   \titlerunning{Hydrodynamic modeling of the accretion shock}

   \maketitle
%

\section{Introduction}

Pre-main sequence stars are strong X-ray emitters, with X-ray luminosity
up to $10^3$ times the solar value ($L_X\sim 10^{27}$ erg s$^{-1}$).
Similarly to main-sequence stars, the X-rays are likely produced by
low density ($n_{\rm e} \sim10^{10}$ cm$^{-3}$) plasma enclosed in
coronal loop structures and heated at temperature $T\sim10^6-10^7$ K
\citep{Feigelson1999ARA&A}. During the last five years, high resolution
($R\sim 600$) X-ray observations of some classical T Tauri stars (CTTSs)
(TW Hya, BP Tau, V4046 Sgr, MP Mus and RU Lupi) have shown the presence of
X-ray plasma at $T\sim 2-3\times 10^6$ K and denser than $10^{11}$ cm$^{-3}$
\citep{Kastner2002ApJ,Schmitt2005A&A,Gunther2006A&A, Argiroffi2007A&A,
Robrade2007A&A}, which suggests an origin different from the coronal one.

\cite{Calvet1998ApJ} and \cite{Lamzin1998ARep} proposed that, X-ray
emission from CTTSs could also be produced by the accreting material.
In fact the material heated by the accretion shock at the base of
the accretion column could reach a temperature $T \sim 10^6$ K.
Starting from the hypothesis of the presence of a stationary strong
accretion shock and solving the conservation laws for the shock-heated
region, \cite{Gunther2007A&A} have quantitatively demonstrated that
the accretion shock model well explains some non coronal features of
the X-ray observations of TW Hya. However, these models are based on
several approximations and do not include a detailed model of the stellar
atmosphere, which indeed determines the shock position and could influence
the profile of pressure and density and, therefore, the thickness of
the shocked region. They assume stationary conditions, but different
studies proved the existence of thermal instabilities in presence of an
accreting flow impacting onto a stellar surface \citep{Langer1981ApJ,
Chevalier1982ApJ, Koldoba2008MNRAS}. Considering that the physical
structure of the shock-heated region and of its temporal evolution
determines the fraction of the accreting energy re-emitted in the X-ray
band and its spectral behavior, further investigations on the issues
discussed above are required to understand better the X-ray emission from
shock-heated plasma, to check their observability and, then, to derive
the physical properties of the accretion process from high resolution
X-ray observations.

Here we address these issues with the aid of a time-dependent hydrodynamic
numerical model describing the impact of an accretion stream onto the
chromosphere of a CTTS. As a first application, we compared the X-ray
spectra of the CTTS MP Mus \citep{Argiroffi2007A&A} with those synthesized
from the results of a simulation tuned on this star.

\section{The model}

We assume that the accretion occurs along a magnetic flux tube linking
the circumstellar disk to the star. We consider the flux tube analogous to
closed coronal loops observed on the Sun; therefore, as done for standard
coronal loop models (e.g. \citealt{Peres1982ApJ, Betta1997A&AS}), we
assume that the plasma moves and transports energy exclusively along the
magnetic field lines. This hypothesis is supported by the low value
of the plasma parameter $\beta = P/(B^2/8\pi) \sim 10^{-2}$, where $P\sim
4\times 10^2$ dyn/cm$^2$ is the expected post-shock zone pressure and
$B \sim 10^3$ G is the typical magnetic field at the surface of a CTTS
(\citealt{Johns-Krull2007ApJ}). Note that plasma density and velocity
are expected to vary across the section of an accretion stream (e.g
\citealt{Romanova2003ApJ}). However, in the case of $\beta \ll 1$,
the stream can be considered as a bundle of ``elementary'' streams, each
characterized by different values of density and velocity. Our
1-D model describes one of these elementary streams.
We limit our analysis to the
impact of the accretion stream onto the chromosphere and consider the
portion of the flux tube close to the star. We assume the magnetic field
lines perpendicular to the stellar surface and a plane parallel geometry.

The impact of the accretion stream onto the chromosphere is modeled by
numerically solving the time-dependent fluid equations of mass, momentum,
and energy conservation, taking into account the gravity stratification,
the thermal conduction (including the effects of heat flux saturation)
and the radiative losses from an optically thin plasma:

\begin{equation}
\frac{\partial \rho}{\partial t} +  \frac{\partial \rho v}{\partial s}
= 0~,
\label{eq:massa-1}
\end{equation}

\begin{equation}
\frac{\partial \rho v}{\partial t} +\frac{\partial (P+\rho v^2)}{\partial
s} = \rho g~,
\label{eq:momento-1}
\end{equation}

\begin{equation}
\frac{\partial \rho E}{\partial t} +\frac{\partial (\rho E+P) v}
{\partial s} = \rho v g + E_{H} -  \frac{\partial q}{\partial s} -
n_e n_H \Lambda(T)~,
\label{eq:en+r+c-1}
\end{equation}

\[
\epsilon =\frac{P}{\rho (\gamma-1)}~,\hspace{0.8cm} P=(1+\beta(\rho,T))\frac{\rho  k_b T}{m_A}
\]

\noindent
where $t$ is the time, $s$ the coordinate along the magnetic field
lines, $\rho$ the mass density, $v$ the plasma velocity, $P$ the
thermal pressure, $g(s)$ the gravity, $T$ the plasma temperature, $E =
\epsilon + v^2/2$ the total gas energy per unit mass, $\epsilon$ the
internal energy per unit mass, $q$ the heat flux in the formulation
of \cite{Spitzer1962} corrected for the effect of flux saturation
\citep{Cowie1977ApJ}, $n_{\rm e}$ and $n_{\rm H}$ are the electron
number density and hydrogen number density, respectively, $\Lambda(T)$
is the radiative losses per unit emission measure, $\beta (\rho,T)$ the
fractional ionization ($n_{\rm e}/n_{\rm H}$) derived from the modified
Saha equation (see \citealt{Brown1973SoPh} for the solar chromosphere
condition), $\gamma$= 5/3 is the ratio between the specific heats, $m_A
= 2.14 \times 10^{-24}$~g is the average atomic mass (assuming heavy
elements abundances 0.5 the solar values; \citealt{Anders1989GeCoA}),
and $E_H$ is a parametrized chromospheric heating function ($E_H =
0$ for $T > 8\times10^3$ K) defined, as in \cite{Peres1982ApJ}, to
keep the unperturbed chromosphere in stable equilibrium. The radiative
losses per unit emission measure are derived by the PINTofALE spectral
code \citep{Kashyap2000BASI} with the APED V1.3 atomic line database
\citep{Smith2001ApJ}, and assuming heavy elements abundances 0.5 the solar
values \citep{Anders1989GeCoA}, as obtained from X-ray observations of
CTTSs by \cite{Telleschi2007A&Ab}. The equations are solved numerically
using the \FLASH\ code (\citealt{Fryxell2000ApJS}), an adaptive mesh
refinement multiphysics code for astrophysical plasmas. The code has been
extended with additional computational modules to handle the radiative
losses, the thermal conduction, and the evolution of fractional ionization
($n_{\rm e}/n_{\rm H}$).

The computational domain extends over a range $D = 4.34 \times
10^9$ cm above the stellar surface. We allow for 5 levels of
refinement in the adaptive mesh algorithm of \FLASH\ (\PARAMESH;
\citealt{MacNeice2000CoPhC}), with resolution increasing twice at each
refinement level: $\Delta s_{max}=2.1 \times 10^6$ cm at the coarsest
resolution, and $\Delta s_{min}=1.3 \times 10^5$ cm at the finest level,
which would correspond to a uniform mesh of $\sim 30000$ grid points. We
analyzed the effect of spatial resolution on our results by considering
two additional simulations which use a setup identical to the one
discussed here, but with either 4 or 6 levels of refinement. We found
that the adopted resolution is the best compromise between accurancy
and computational cost and that the system evolution is well described
in its detail.

The simulation presented here covers a time interval of about 2000 s.
We used the accretion parameters (velocity and density) derived
by \cite{Argiroffi2007A&A} to match the soft X-ray emission of MP Mus.
We calculated the gravity considering the star mass $M=1.2 M_{\sun}$
and the star radius $R=1.3 R_{\sun}$ used by \cite{Argiroffi2007A&A}.

The external part of the initial configuration, extending from $2.75
\times 10^8$ to $4 \times 10^9$ cm, consists of an accretion stream
constant in density ($n_{\rm e}=10^{11}$ cm$^{-3}$), temperature ($T=
10^3$ K) and velocity ($v=450$ km/s). The inner part of the initial
configuration consists of a static chromosphere. We reproduce the pressure
gradient of a young stellar chromosphere by considering the temperature
profiles prescribed by \cite{Vernazza1973ApJ} solar chromosphere models
scaled in order to match a pressure of $\sim 7\times 10^4$ dyn cm$^{-2}$
at the base of the chromosphere. As boundary conditions we consider fixed
values both at the top ($n_{\rm acc}=10^{11}$ cm$^{-3}$, $T_{\rm acc}=
10^3$ K, $v_{\rm acc}=450$ km/s) and at the base ($n_{\rm chr} =10^{17}$
cm$^{-3}$, $T_{\rm chr} =4.44 \times 10^3$ K, $v_{\rm chr} =0.0$ km/s)
of the computational domain. In principle these boundary conditions
lead to accumulation of matter at the base of the chromosphere. However,
we have estimated that this effect becomes significant on a timescale 200
times longer than that explored by our simulation. In fact, we checked
that, for $s < 10^8$ cm, the chromosphere remains virtually unperturbed
(with variations of mass density below 1\%) during the timescale
considered. Note also that we neglect the heating of the chromosphere (in particular,
at the lower boundary) due to the X-ray emission originating from hot plasma. 
In the case of MP Mus (effective temperature
$T_{\rm eff} \sim 5000$ K), this approximation is justified by the
low ratio between the energy flux coming from the accretion flow
and from the photospheric emission, $(\rho_{\rm acc}v_{\rm
acc}^3)/(4\sigma T_{\rm eff}^4) \approx 0.1$, where $\sigma$ is the
Stefan-Boltzmann constant. The stability of the chromosphere was tested by
dedicated simulations longer than 100 ks, some of which considering also
strong transient heating.

\section{Results}
\label{results}

\begin{figure}
\includegraphics[width=9 cm]{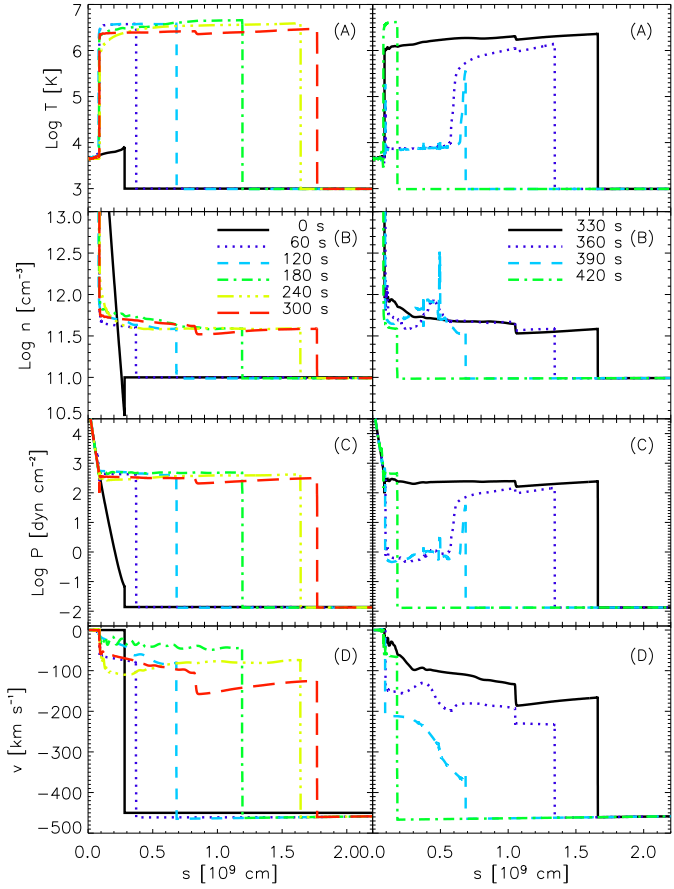}
\caption{Evolution of plasma temperature (A), density (B), pressure
         (C), and velocity (D) distributions along the flux tube from
         the chromosphere to the unperturbed accretion stream, sampled
         every 60 s from 0 to 300 s (left panels) and every 30 s from
         330 to 420 s (right panels). The figure shows the inner portion
         of the spatial domain, including the chromosphere and the
         hot slab.}
\label{fig:sim_result} \end{figure}

The impact of the accretion stream onto the stellar chromosphere leads
to the formation of transmitted (into the chromosphere) and reverse (into
the accretion column) shocks. The latter propagates through the accretion
column building up an hot slab. As pointed out by \cite{Argiroffi2007A&A},
in the strong shock limit \citep{ZelDovich1967book}, the expected
temperature of the slab is $\sim(3/16)(\mu m_{\rm H}/k_{\rm
b})v_{acc}^2\approx 3\times 10^6$ K and, since it is subject to the
radiative cooling, the expected maximum thickness is $\sim v_{\rm
ps}\tau_{rad}$, where $v_{\rm ps} = v_{\rm acc}/4 \approx 100$~km s$^{-1}$
is the post-shock plasma velocity in the slab and $\tau_{rad}$ is the
cooling time for the shocked gas defined as

\begin{equation}
\tau_{\rm rad}=\frac{1}{\gamma-1}\frac{P}{n_{\rm e}n_{\rm H}\Lambda(T)}\sim
6.7\times10^{3}\frac{T^{3/2}}{n_{\rm e}}~,
\end{equation}

\noindent
where, for temperatures characteristic of the slab ($T \approx 10^5-10^7$
K), we approximated the cooling function as $\Lambda(T) \approx 6.0\times
10^{-20} T^{-1/2}$ erg s$^{-1}$ cm$^{3}$. For the typical values of $T$
and $n_{\rm e}$ of the hot slab ($T\approx 3\times 10^6$~K and $n_{\rm e}
= 4\,n_{\rm acc} \sim 4\times 10^{11}$~cm$^{-3}$), $\tau_{\rm rad}\approx
90$~s and the thickness of the slab is expected to be $D_{\rm slab}
\approx 10^9$~cm.  Note that deviations from equilibrium of ionization
may be present beyond the shock surface due to the rapid growth of
temperature (e.g. \citealt{Lamzin1998ARep}). However, these deviations
are expected within a distance $d_{\rm NEI} \la v_{\rm ps}\tau_{eq}\sim
10^7$ cm from the shock, where $\tau_{eq}$, the ionization equilibrium
timescale, has been estimated for the characteristic density and
temperature of the slab. Being $d_{\rm NEI} \ll D_{\rm slab}$, the
non-equilibrium ionization effects are not relevant for the evolution
of the hot slab.

The evolution of temperature, density, pressure and velocity during the
first part of our simulation is shown in Fig. \ref{fig:sim_result}.
During the first 300~s of evolution, the reverse shock gradually
heats up the accretion column at $T\sim 3\times 10^6$~K and the
inflow piles up material creating a hot slab with density $n_{\rm e}
= 3-7\times$10$^{11}$ cm$^{-3}$ and constant pressure equal to the
accretion flow ram pressure $P_{\rm ram}= \rho_{\rm acc} v_{\rm acc}^2
= 450$ dyn cm$^{-2}$ (see left panels of Fig.~\ref{fig:sim_result}).
The base of the shocked region initially penetrates the chromosphere
and comes at rest where the ram pressure is equal to the thermal pressure. The
ram pressure, therefore, determines the position of the slab inside the
chromosphere, while the chromosphere below the shock region remains
virtually unperturbed. Due to pile up of material at the base of the
shocked column, the radiative losses gradually increase there. At the
end of this phase, the extension of the hot slab is $\approx 1.8\times
10^9$~cm.

In the subsequent phase of evolution, the strong radiative cooling
of the high density plasma accumulated at the base of the hot slab
triggers thermal instabilities there (\citealt{Field1965ApJ}; see right panels
of Fig.~\ref{fig:sim_result}): plasma temperature and pressure decrease
by more than two orders of magnitude in few seconds, leading to the
collapse of the upper layers of the accretion column. As a consequence,
the residual hot slab fells down and the reverse shock moves downward to
the chromosphere. The additional compression due to the collapse leads
to the further increase of plasma density and radiative cooling at the
base of the hot slab which gradually cools down in $\sim 100$~s.

After the accretion column is completely cooled, a new hot slab
is generated (see the dashed-dotted lines in the right panels of
Fig. \ref{fig:sim_result}) and the system starts a quasi-periodic
evolution with alternating heating and cooling phases lasting $\sim$
400 s linked to the up and down displacement of the reverse shock. Our
simulation tracks the system evolution for 5 periods without significant
differences among them. Therefore, we presume that the results can
be extended to longer time intervals. Note that the period of the
fluctuation depends on the parameters of the inflow and partially of
the chromosphere. The assumption of steady flow is appropriate over
the limited time explored. The whole issue should be revisited for
simulations covering timescales comparable to those of changes due to
stellar rotation, magnetic field evolution, etc. .

Note that \cite{Langer1981ApJ} have found periodic variations of the
accretion shock position and, therefore, of the thickness of the hot slab
in the context of compact objects. As in our case, the cycle consists
of time intervals when the shock moves upward and the gas accumulates
behind it and phases during which the shock moves downward as the hot gas
cools radiatively. In the context of CTTSs, \cite{Koldoba2008MNRAS} found
oscillations of the accretion shock position with periods $\sim 0.02-0.2$,
due to the same mechanism described here.
These short periods are mainly due to the set of values of accretion
flow parameters (velocity and density) and heavy elements abundances
adopted by these authors. In addition, our model includes the effects
of thermal conduction (important in the energy budget) and the stellar
chromosphere (which determines the shock position).

We checked the observability of the X-ray emission produced by the
shock-heated material by synthesizing the spectrum in the energy
range $[0.5-8.0]$ KeV from the simulation results, and by calculating
the time-average luminosity over the time interval covered by the
simulation\footnote{High resolution X-ray spectra of CTTSs are gathered
with exposure times of $\sim 10-100$~ks, i.e. much larger than the
time interval covered by our simulation.} Since our model is 1-D,
we have to assume the flux tube cross section. Considering the velocity
of the accretion stream $v=450$ km s$^{-1}$, its density $n_{\rm
e}=10^{11}$~cm$^{-3}$ and an accretion rate $\sim 10^{-10}~M_{\sun}$
yr$^{-1}$, the accretion stream cross section is $\sim 6.5\times 10^{20}$
cm$^2$. Using this cross section, we obtain an X-ray luminosity varying from $\sim 8.4\times 10^{21}$  to
$\sim1.6 \times 10^{30}$ erg s$^{-1}$ with a time-average value of $L_{\rm
X} = 7.2 \times 10^{29}$ erg s$^{-1}$. Since this value is comparable
with the typical overall luminosity observed in the young T Tauri stars
of the same mass (\citealt{Preibisch2005ApJS}), our model shows that the
shock-heated material can contribute to the X-ray emission of the CTTSs.

\begin{figure}
\centering
\includegraphics[width=9. cm]{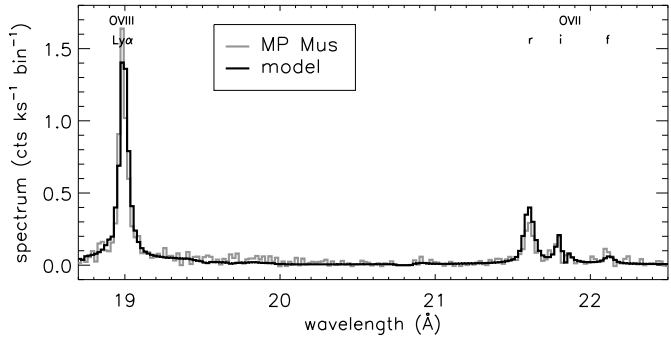}
\caption{Observed X-ray spectrum of the star MP Mus (gray line) with
          the synthetic spectrum derived from the simulation
         (black line).}
\label{fig:spectra}
\end{figure}

In Fig.~\ref{fig:spectra}, we compare the X-ray spectrum synthesized
from the simulation (averaged over the time interval covered by
the simulation) with the spectrum of the star MP Mus observed with the
Reflection Grating Spectrometers (RGS) on board the XMM-Newton satellite
\citep{Argiroffi2007A&A}. The synthesis of the X-ray spectrum takes into
account the instrumental response of XMM-RGS and the interstellar
absorption ($N_{\rm H} =5\times 10^{20}$ cm$^{-2}$) derived from the
observations, and assumes the emitting source at the same distance of MP
Mus (86 pc). We also assumed that the time-average X-ray luminosity
produced by the shock-heated slab in the $18-23$ {\AA} wavelength range
is equal to the observed MP Mus luminosity in the same range. With
the above assumptions, it turns out that the accretion rate is $\sim
7.7\times 10^{-11}~M_{\sun}$ yr$^{-1}$ and the accretion stream cross
section is $\sim 5\times 10^{20}$ cm$^2$ (i.e. a surface filling factor
of about 0.5\%). These values are in agreement with those ($\sim5\times 
10^{-11}~M_{\sun}$ yr$^{-1}$ and $\sim 3\times 10^{20}$ cm$^2$) derived 
by \cite{Argiroffi2007A&A} from the analysis of the observations. 

Figure \ref{fig:spectra} shows \ion{O}{viii} Ly$\alpha$ and the \ion{O}{vii} triplet. 
The ratio between the \ion{O}{viii} Ly$\alpha$ and the \ion{O}{vii}
resonance line is a tracer of plasma temperature and the ratio between
the \ion{O}{vii} forbidden and intercombination lines is a tracer of
plasma density. The good agreement
between the observed and the synthetic line profiles demonstrates that
our model is able to explain the origin of the whole dense X-ray plasma
observed in CTTSs.  Furthermore, our hypothesis on heavy elements
abundances is supported by the agreement of the ratio between lines
and continuum intensity.

\section{Conclusions}

We model the impact of an accretion flow on the chromosphere of a young
T Tauri star.  Our main results are:

\begin{itemize}
\item the impact is able to build up a slab of accreting plasma
with density $n_{\rm e} \sim 5\times 10^{11}$ cm$^{-3}$ and temperature
$T\sim 3\times 10^6$ K;

\item the pressure along the slab is uniform and equal to the ram pressure
of the accretion shock; therefore, the slab is rooted down in the chromosphere
where the thermal pressure is equal to the ram pressure;

\item the thickness of the slab grows until it reaches $\sim 1.8\times
10^9$ cm, when it undergoes a thermal collapse due to the enhancement
of density and, therefore, of radiative losses at its base;

\item the system oscillates quasi-periodically between the formation of
an hot slab and its thermal collapse by thermal instability;

\item according to our model, an accretion rate of $\sim 10^{-10}$
$M_{\sun}$ yr$^{-1}$ is enough to produce an X-ray
luminosity varying between $\sim 8.4\times 10^{21}$ and $1.6\times 10^{30}$ erg $s^{-1}$, with 
a time-average value of $L_{\rm X} \sim 7 \times 10^{29}$ erg s$^{-1}$, which is
comparable to the overall X-ray luminosity observed in the CTTSs
of the same mass;

\item our model is able to explain in detail the observed X-ray spectrum
of the star MP Mus \citep{Argiroffi2007A&A}.

\end{itemize}

A large set of simulations, exploring the domain of the physical parameters
of the system, will be studied in a future paper.

\begin{acknowledgements}
We thank Jeremy Drake for useful discussions. This work was supported
in part by the Italian Ministry of University and Research (MIUR)
and by Istituto Nazionale di Astrofisica (INAF). The software used in
this work was in part developed by the DOE-supported ASC / Alliance
Center for Astrophysical Thermonuclear Flashes at the University of
Chicago. This work makes use of results produced by the PI2S2 Project
managed by the Consorzio COMETA, a project co-funded by the Italian
Ministry of University and Research (MIUR) within the Piano Operativo
Nazionale ``Ricerca Scientifica, Sviluppo Tecnologico, Alta Formazione''
(PON 2000-2006). More information is available at http://www.pi2s2.it
and http://www.consorzio-cometa.it.
\end{acknowledgements}

\bibliographystyle{aa}
\bibliography{biblio}

\begin{thebibliography}{29}
\expandafter\ifx\csname natexlab\endcsname\relax\def\natexlab#1{#1}\fi

\bibitem[{{Anders} \& {Grevesse}(1989)}]{Anders1989GeCoA}
{Anders}, E. \& {Grevesse}, N. 1989, \gca, 53, 197

\bibitem[{{Argiroffi} {et~al.}(2007){Argiroffi}, {Maggio}, \&
  {Peres}}]{Argiroffi2007A&A}
{Argiroffi}, C., {Maggio}, A., \& {Peres}, G. 2007, \aap, 465, L5

\bibitem[{{Betta} {et~al.}(1997){Betta}, {Peres}, {Reale}, \&
  {Serio}}]{Betta1997A&AS}
{Betta}, R., {Peres}, G., {Reale}, F., \& {Serio}, S. 1997, \aaps, 122, 585

\bibitem[{{Brown}(1973)}]{Brown1973SoPh}
{Brown}, J.~C. 1973, \solphys, 29, 421

\bibitem[{{Calvet} \& {Gullbring}(1998)}]{Calvet1998ApJ}
{Calvet}, N. \& {Gullbring}, E. 1998, \apj, 509, 802

\bibitem[{{Chevalier} \& {Imamura}(1982)}]{Chevalier1982ApJ}
{Chevalier}, R.~A. \& {Imamura}, J.~N. 1982, \apj, 261, 543

\bibitem[{{Cowie} \& {McKee}(1977)}]{Cowie1977ApJ}
{Cowie}, L.~L. \& {McKee}, C.~F. 1977, \apj, 211, 135

\bibitem[{{Feigelson} \& {Montmerle}(1999)}]{Feigelson1999ARA&A}
{Feigelson}, E.~D. \& {Montmerle}, T. 1999, \araa, 37, 363

\bibitem[{{Field}(1965)}]{Field1965ApJ}
{Field}, G.~B. 1965, \apj, 142, 531

\bibitem[{{Fryxell} {et~al.}(2000){Fryxell}, {Olson}, {Ricker}, {Timmes},
  {Zingale}, {Lamb}, {MacNeice}, {Rosner}, {Truran}, \&
  {Tufo}}]{Fryxell2000ApJS}
{Fryxell}, B., {Olson}, K., {Ricker}, P., {et~al.} 2000, \apjs, 131, 273

\bibitem[{{G{\"u}nther} {et~al.}(2006){G{\"u}nther}, {Liefke}, {Schmitt},
  {Robrade}, \& {Ness}}]{Gunther2006A&A}
{G{\"u}nther}, H.~M., {Liefke}, C., {Schmitt}, J.~H.~M.~M., {Robrade}, J., \&
  {Ness}, J.-U. 2006, \aap, 459, L29

\bibitem[{{G{\"u}nther} {et~al.}(2007){G{\"u}nther}, {Schmitt}, {Robrade}, \&
  {Liefke}}]{Gunther2007A&A}
{G{\"u}nther}, H.~M., {Schmitt}, J.~H.~M.~M., {Robrade}, J., \& {Liefke}, C.
  2007, \aap, 466, 1111

\bibitem[{{Johns-Krull}(2007)}]{Johns-Krull2007ApJ}
{Johns-Krull}, C.~M. 2007, \apj, 664, 975

\bibitem[{{Kashyap} \& {Drake}(2000)}]{Kashyap2000BASI}
{Kashyap}, V. \& {Drake}, J.~J. 2000, Bulletin of the Astronomical Society of
  India, 28, 475

\bibitem[{{Kastner} {et~al.}(2002){Kastner}, {Huenemoerder}, {Schulz},
  {Canizares}, \& {Weintraub}}]{Kastner2002ApJ}
{Kastner}, J.~H., {Huenemoerder}, D.~P., {Schulz}, N.~S., {Canizares}, C.~R.,
  \& {Weintraub}, D.~A. 2002, \apj, 567, 434

\bibitem[{{Koldoba} {et~al.}(2008){Koldoba}, {Ustyugova}, {Romanova}, \&
  {Lovelace}}]{Koldoba2008MNRAS}
{Koldoba}, A.~V., {Ustyugova}, G.~V., {Romanova}, M.~M., \& {Lovelace},
  R.~V.~E. 2008, \mnras, 388, 357

\bibitem[{{Lamzin}(1998)}]{Lamzin1998ARep}
{Lamzin}, S.~A. 1998, Astronomy Reports, 42, 322

\bibitem[{{Langer} {et~al.}(1981){Langer}, {Chanmugam}, \&
  {Shaviv}}]{Langer1981ApJ}
{Langer}, S.~H., {Chanmugam}, G., \& {Shaviv}, G. 1981, \apjl, 245, L23

\bibitem[{{MacNeice} {et~al.}(2000){MacNeice}, {Olson}, {Mobarry}, {de
  Fainchtein}, \& {Packer}}]{MacNeice2000CoPhC}
{MacNeice}, P., {Olson}, K.~M., {Mobarry}, C., {de Fainchtein}, R., \&
  {Packer}, C. 2000, Computer Physics Communications, 126, 330

\bibitem[{{Peres} {et~al.}(1982){Peres}, {Serio}, {Vaiana}, \&
  {Rosner}}]{Peres1982ApJ}
{Peres}, G., {Serio}, S., {Vaiana}, G.~S., \& {Rosner}, R. 1982, \apj, 252, 791

\bibitem[{{Preibisch} {et~al.}(2005){Preibisch}, {Kim}, {Favata}, {Feigelson},
  {Flaccomio}, {Getman}, {Micela}, {Sciortino}, {Stassun}, {Stelzer}, \&
  {Zinnecker}}]{Preibisch2005ApJS}
{Preibisch}, T., {Kim}, Y.-C., {Favata}, F., {et~al.} 2005, \apjs, 160, 401

\bibitem[{{Robrade} \& {Schmitt}(2007)}]{Robrade2007A&A}
{Robrade}, J. \& {Schmitt}, J.~H.~M.~M. 2007, \aap, 473, 229

\bibitem[{{Romanova} {et~al.}(2003){Romanova}, {Ustyugova}, {Koldoba}, {Wick},
  \& {Lovelace}}]{Romanova2003ApJ}
{Romanova}, M.~M., {Ustyugova}, G.~V., {Koldoba}, A.~V., {Wick}, J.~V., \&
  {Lovelace}, R.~V.~E. 2003, \apj, 595, 1009

\bibitem[{{Schmitt} {et~al.}(2005){Schmitt}, {Robrade}, {Ness}, {Favata}, \&
  {Stelzer}}]{Schmitt2005A&A}
{Schmitt}, J.~H.~M.~M., {Robrade}, J., {Ness}, J.-U., {Favata}, F., \&
  {Stelzer}, B. 2005, \aap, 432, L35

\bibitem[{{Smith} {et~al.}(2001){Smith}, {Brickhouse}, {Liedahl}, \&
  {Raymond}}]{Smith2001ApJ}
{Smith}, R.~K., {Brickhouse}, N.~S., {Liedahl}, D.~A., \& {Raymond}, J.~C.
  2001, \apjl, 556, L91

\bibitem[{{Spitzer}(1962)}]{Spitzer1962}
{Spitzer}, L. 1962, {Physics of Fully Ionized Gases} (New York: Wiley)

\bibitem[{{Telleschi} {et~al.}(2007){Telleschi}, {G{\"u}del}, {Briggs},
  {Audard}, \& {Scelsi}}]{Telleschi2007A&Ab}
{Telleschi}, A., {G{\"u}del}, M., {Briggs}, K.~R., {Audard}, M., \& {Scelsi},
  L. 2007, \aap, 468, 443

\bibitem[{{Vernazza} {et~al.}(1973){Vernazza}, {Avrett}, \&
  {Loeser}}]{Vernazza1973ApJ}
{Vernazza}, J.~E., {Avrett}, E.~H., \& {Loeser}, R. 1973, \apj, 184, 605

\bibitem[{{Zel'Dovich} \& {Raizer}(1967)}]{ZelDovich1967book}
{Zel'Dovich}, Y.~B. \& {Raizer}, Y.~P. 1967, {Physics of shock waves and
  high-temperature hydrodynamic phenomena, ed. Hayes, W.~D., Probstein, R.~F.}
  (New York: Academic Press)

\end{thebibliography}

\end{document}